# Vision-aided nonlinear control framework for shake table tests

Zhongwei Chen[1*], T.Y. Yang[2], Yifei Xiao[3], Xiao Pan[4], WanyanYang[5]

[1] Master student, Department of Civil Engineering, University of British Columbia, abchenzw@mail.ubc.ca
[2] Professor, Department of Civil Engineering, University of British Columbia
[3] Ph.D. student, Department of Civil Engineering, University of British Columbia
[4] Research fellow, Department of Civil Engineering, University of British Columbia
[5] Undergraduate intern, Department of Civil Engineering, Harbin Institute of Technology

**ABSTRACT**

The structural response under the earthquake excitations can be simulated by scaled-down model shake table tests or full-scale model shake table tests. In this paper, adaptive control theory is used as a nonlinear shake table control algorithm which considers the inherent nonlinearity of the shake table system and the Control-Structural Interaction (CSI) effect that the linear controller cannot consider, such as the Proportional-Integral-Derivative (PID) controller. The mass of the specimen can be assumed as an unknown variation and the unknown parameter will be replaced by an estimated value in the proposed control framework. The signal generated by the control law of the adaptive control method will be implemented by a loop-shaping controller. To verify the stability and feasibility of the proposed control framework, a simulation of a bare shake table and experiments with a bare shake table with a two-story frame were carried out. This study randomly selects Earthquake recordings from the Pacific Earthquake Engineering Research Center (PEER) database. The simulation and experimental results show that the proposed control framework can be effectively used in shake table control.

## 1. INTRODUCTION

In earthquake engineering, shake table control is an important issue. The main difficulty for the shake table control is that the upper structure's dynamic behaviour will affect the shake table's dynamic behaviour. While many passive control methods have been widely adopted in earthquake engineering applications (e.g., [1-2]), the main control algorithms for the shake table test are active control methods such as the PID control and three-variable control [23-24]. However, these two controllers are linear controllers, which cannot consider the CSI effect. A series of nonlinear controllers have been proposed recently to compensate for the CSI effect and improve the tracking performance in displacement, velocity, and acceleration [3-4].

Enokida et al [5] proposed the nonlinear signal-based control framework for a uniaxial shake table with a nonlinear upper structure, which can be designed by using the classical control theory and expressed as the form of the transfer function. The proposed control framework was verified by both simulation and experiments. The results show the output of nonlinear signal-based control can achieve nearly 100% accuracy. Yang et al [6] proposed a Lyapunov-based nonlinear control algorithm and the shake table test with a one-sixth-scaled model was carried out the prove the performance and stability of the proposed control framework. The results present that the proposed control method is not only a viable alternative but also improves the quality of the shake table tests. Xiao et al [7] proposed a nonlinear backstepping control algorithm with a high-gain observer used to provide accurate feedback signals.

Adaptive control has been proposed and used in the shake table control [8-10]. The main purpose of adaptive control is to maintain the consistent performance of a system in the presence of unknown variations. Adaptive control has two main methods, one is the model reference adaptive control, and the other is self-tuning adaptive control. In this paper, the model reference adaptive control is adopted. The mass of the upper structure is assumed as an unknown parameter. The proposed control framework is verified by both simulation and experiments.

Since the proposed control framework requires displacement, velocity, and acceleration feedback, the traditional method is to use a linear pot and LVDT to measure the displacement of the upper structure and the shake table. The displacement of the upper structure and shake table can also be measured by using computer vision methods (e.g., [11]). More details will be discussed in Section 5.

## 2. METHODOLOGY

The final control law will be obtained in this section based on the adaptive control theory and the physical model of a uniaxial shake table model with a two-degrees of freedom system.

### 2.1. Dynamic equilibrium equation and state space equation

The physical model is shown in Figure (1). The dynamics equilibrium equation can be obtained and shown in Equation (1).

$$F = m_t \ddot{d}_t + m_1 \ddot{d}_1 + m_2 \ddot{d}_2 \tag{1}$$

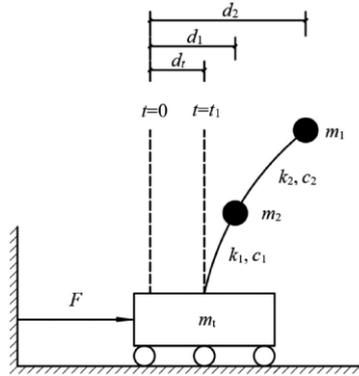

Figure 1 Physical model of the shake table with a two degree of freedom system [7]

where $F$ is the force from the actuator. $d_t, d_1, d_2$ are the displacement of the shake table, the first level of the upper structure and the second level of the upper structure respectively. $\dot{d}_t, \dot{d}_1, \dot{d}_2$ are the velocity of shake table, the first level of the upper structure and second level of the upper structure. $\ddot{d}_t, \ddot{d}_1, \ddot{d}_2$ are the acceleration of shake table, the first level of the upper structure and second level of the upper structure. $m_t, m_1, m_2$ are the mass of the shake table, the first level of the upper structure and the second level of the upper structure. $k_1$ and $k_2$ are the stiffness of the first and second floors. $c_1$ and $c_2$ are the damping ratios of the first and second floors.

The dynamic equilibrium equation can be rewritten as the state space equation form, which is shown in Equation (2).

$$\boldsymbol{Xp} = \begin{Bmatrix} x_{p1} \\ x_{p2} \end{Bmatrix} = \begin{Bmatrix} d_t \\ \dot{d}_t \end{Bmatrix}, \dot{\boldsymbol{Xp}} = \begin{Bmatrix} \dot{x}_{p1} \\ \dot{x}_{p2} \end{Bmatrix} = \begin{Bmatrix} \dot{d}_t \\ \ddot{d}_t \end{Bmatrix} = \begin{Bmatrix} x_{p2} \\ \frac{1}{m_t}F - \frac{m_1}{m_t}\ddot{d}_1 - \frac{m_2}{m_t}\ddot{d}_2 \end{Bmatrix} \tag{2}$$

### 2.2. Derivation of the control law

The augmented matrix $\boldsymbol{X}(t) = \begin{bmatrix} \boldsymbol{X}_p^T(t) & \boldsymbol{X}_c^T(t) \end{bmatrix}^T$ (the dimension of $\boldsymbol{X}(t)$ is is valued $R^a$) is shown in Equation (3).

$$\dot{\boldsymbol{X}}(t) = \boldsymbol{A} \cdot \boldsymbol{X}(t) + \boldsymbol{B} \cdot \wedge \cdot F + \boldsymbol{B} \cdot \boldsymbol{Wp}^T \cdot \boldsymbol{\sigma p}(t) + \boldsymbol{B_r} \cdot \boldsymbol{c}(t) \tag{3}$$

where $A = \begin{bmatrix} A_p & 0_{nXn_c} \\ E_p & E_r \end{bmatrix}, B = \begin{bmatrix} B_p \\ 0_{n_cXm} \end{bmatrix}, B_r = \begin{bmatrix} 0_{nXn_c} \\ -I_{n_cXn_c} \end{bmatrix}$, $A_p = \begin{bmatrix} 0 & 1 \\ 0 & 0 \end{bmatrix}$ $B_p = \begin{bmatrix} 0 \\ 1 \end{bmatrix}$ $\wedge = \begin{bmatrix} \frac{1}{m_t} \end{bmatrix}$ $Wp^T = \begin{bmatrix} -\frac{m_1}{m_t} & -\frac{m_2}{m_t} \end{bmatrix}$ $\sigma p(t) = \begin{bmatrix} \ddot{d}_1 \\ \ddot{d}_2 \end{bmatrix}$. $E_p$ is matrix $0^{n_cXn}$. $E_r$ is matrix $I^{n_cXn_c}$ [12]. $X_c$ is shown in Equation (4).

$$\dot{X}_c(t) = E_r \cdot X_c(t) - c(t) \tag{4}$$

where $c(t)$ is shown in Equation (5).

$$c(t) = E_r \cdot r(t) - \dot{r}(t) \tag{5}$$

where $r$ is the matrix of reference input valued $R^{n_c}$.

Equation (3) can be rewritten as Equation (6).

$$\dot{X}(t) = A_r \cdot X(t) + B_r \cdot c(t) + B \cdot \wedge \cdot [-K \cdot X(t) + F + \wedge^{-1} \cdot Wp^T \cdot \sigma p(t)] \tag{6}$$

where $K = [K_p \quad K_r]$. $K$ is valued $R^{m X a}$. $A = A_r - B \cdot K$. $K_p$ and $K_r$ is a constant matrix with dimension of $R^{mXn_c}$. $K_p$ and $K_r$ should be selected such that the eigenvalue of $A_r$ have the strictly negative real part. Define $W^T = [\wedge^{-1} \cdot Wp^T \quad I]$, $\emptyset = [\sigma p(t) \quad -K \cdot X(t)]^T$, where $W^T$ include an unknown parameter, $\emptyset$ is a known parameter. So, Equation (6) can be rewritten as Equation (7).

$$\dot{X}(t) = A_r \cdot X(t) + B_r \cdot c(t) + B \cdot \wedge \cdot [F + W^T \cdot \emptyset] \tag{7}$$

The equation of the reference model can be expressed as Equation (8)

$$\dot{X}_r(t) = A_r \cdot X_r(t) + B_r \cdot c(t) \tag{8}$$

where $X_r(t) = [X_{pr}^T(t) \quad r^T]^T$.

Based on Equation (7) and Equation (8), the response of the shake table test under the earthquake excitation will track the reference trajectory provided by the reference model when the adaptive control law is $F = -W^T \cdot \emptyset$.

Since $W^T$ is an unknown parameter, so an estimated value $\widehat{W}^T$ is used to replace the unknown parameter. Hence the adaptive control law will be expressed as $F = -\widehat{W}^T \cdot \emptyset$, and the error between the actual value and the estimated value can be expressed as $\widetilde{W} = \widehat{W} - W$. Equation (7) can be rewritten as Equation (9).

$$\dot{X}(t) = A_r \cdot X(t) + B_r \cdot c(t) - B \cdot \wedge \cdot \widetilde{W}^T \cdot \emptyset \tag{9}$$

$e = X(t) - X_r(t)$ is defined as the error between the control system and the reference model, and the time derivative of $e$ can be expressed as Equation (10).

$$\dot{e} = A_r \cdot e - B \cdot \wedge \cdot \widetilde{W}^T \cdot \emptyset \tag{10}$$

The stability of the proposed control algorithm should be proved by the Lyapunov theory. The Lyapunov function candidate can be expressed as Equation (11), which is a positive definition function.

$$V = e^T \cdot P \cdot e + \gamma^{-1} tr[\left(\widetilde{W}\wedge^{\frac{1}{2}}\right)^T \cdot \left(\widetilde{W}\wedge^{\frac{1}{2}}\right)] \tag{11}$$

where $P$ is a positive-definite matrix valued $R^{aXa}$, and $\gamma$ is a tuning parameter that is a positive constant. The time derivate of the Lyapunov function can be expressed as Equation (12).

$$\dot{V} = 2 \cdot e^T \cdot P \cdot \dot{e} + 2 \cdot \gamma^{-1} \cdot tr\left(\wedge \cdot \widetilde{W}^T \cdot \dot{\widehat{W}}\right) \tag{12}$$

Equation (12) can be rewritten as Equation (13).

$$\dot{V} = e^T \cdot \left(A_r^T \cdot P + P \cdot A_r\right) \cdot e - 2 \cdot \gamma^{-1} tr\left(\wedge \cdot \widetilde{W}^T \cdot \left(\gamma \cdot \emptyset \cdot e^T \cdot P \cdot B - \dot{\widehat{W}}\right)\right) \tag{13}$$

where $A_r^T \cdot P + P \cdot A_r = -I$ is the Lyapunov equation.

To satisfy the requirement of stability, the time derivate of the Lyapunov function candidate should be a negative definition, which ensures the error will approach 0 asymptotically. Hence, $\gamma \cdot \emptyset \cdot e^T \cdot P \cdot B - \hat{W}$ should be equal to 0. Therefore, the control signal can be expressed as Equation (14).

$$F = -\int [\gamma \cdot \emptyset \cdot e^T \cdot P \cdot B]^T dt \cdot \emptyset \tag{14}$$

### 2.3. Control framework

Figure (2) presents the proposed control framework, the control signal to the loop-shaping controller is a displacement command, and the filter used in this paper is the second-order low-pass Butterworth filter with a cut-off frequency of 50 Hz.

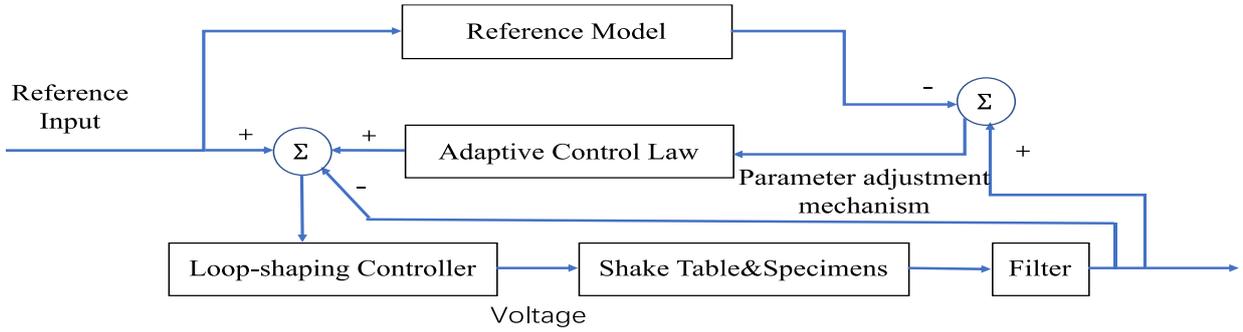

Figure 2 Block diagram of the proposed control framework

### 2.4. Vision-based vibration measurement

Computer vision has been recognized to provide effective solutions in various structural engineering applications such as damage detection of various structural systems and components [14-21], and structural vibration measurement [11, 22]. The proposed control framework requires structural vibration response from multiple upper floors. Therefore, it will be efficient to apply computer vision methods to provide contactless sensing feedback for the upper floors when executing the shake table tests. In this study, the vision-based vibration method [11] is adopted to provide sensor feedback for the control framework.

## 3. NUMERICAL SIMULATION AND EXPERIMENTS

### 3.1. The transfer function of the shake table

The transfer function of the hydraulic shake table system can be obtained by system identification. Equation (15) and Equation (16) shows the transfer function from servo valve voltage to the shake table displacement and acceleration.

$$G_{v-d} = \frac{719s^5 + 3.13e6s^4 + 6.73e9s^3 + 7.83e13s^2 + 3.95e15s}{s^8 + 309s^7 + 1.67e5s^6 + 3.2e7s^5 + 6.77e9s^4 + 7.5e11s^3 + 5.98e13s^2} \tag{15}$$

$$G_{v-a} = \frac{719s^5 + 3.13e6s^4 + 6.73e9s^3 + 7.83e13s^2 + 3.95e15s}{s^6 + 309s^5 + 1.67e5s^4 + 3.2e7s^3 + 6.77e9s^2 + 7.5e11s^1 + 5.98e13} \tag{16}$$

## 3.2. Simulation

The simulation evaluation analysis was performed in MATLAB/Simulink, where the transfer function of the bare shake table was adopted to simulate the hydraulic shake table system. Two ground motions selected randomly from the Pacific Earthquake Engineering Research Center database [13] were used: RSN 4013 and RSN 1006, respectively. The acceleration time series curve is shown in Figure (2).

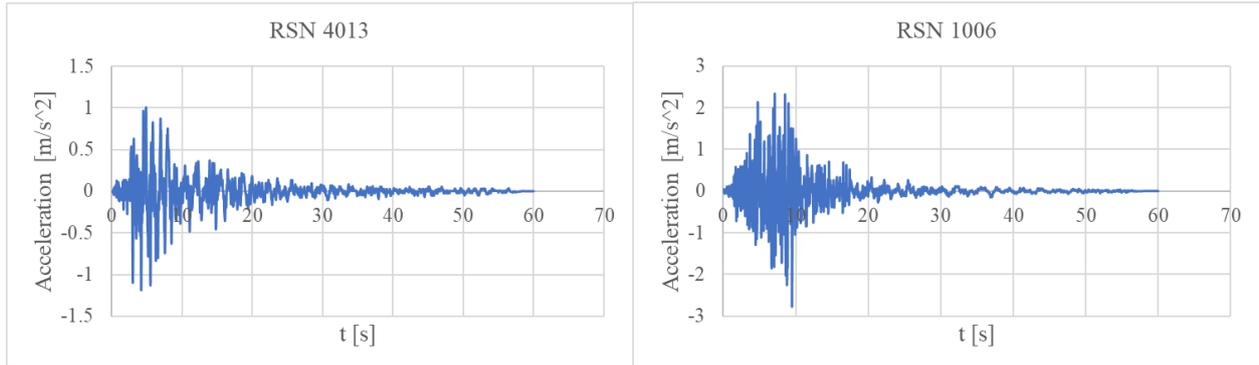

Figure 3 Time series curve of the selected ground motions

The comparison between the reference signal and the simulation results from MATLAB/Simulink is shown in Figure (4) and Figure (5), including the displacement, velocity, and acceleration results. The simulation results present that the proposed control framework is stable, and the simulated output of the shake table can reach the desired results.

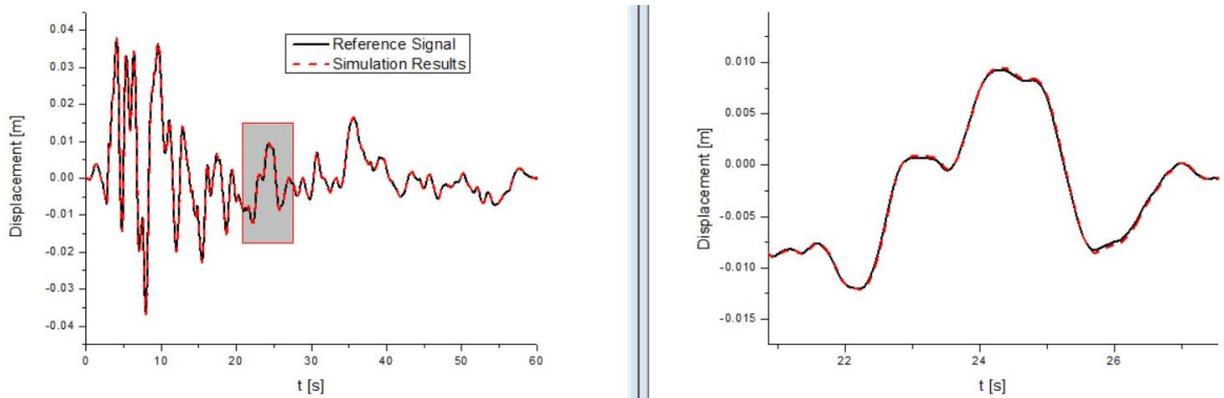

*Figure 4a Displacement results of RSN 4013 from simulation*

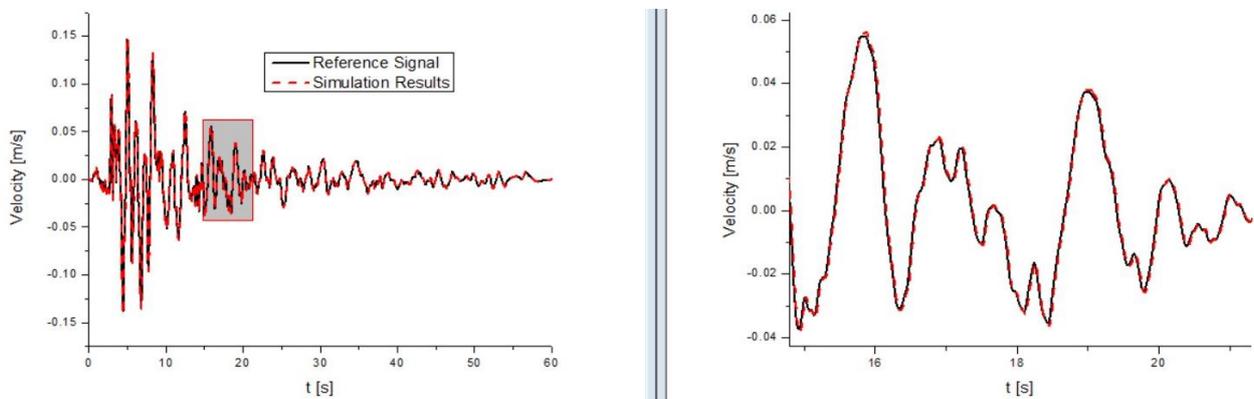

*Figure 4b Velocity results of RSN 4013 from simulation*

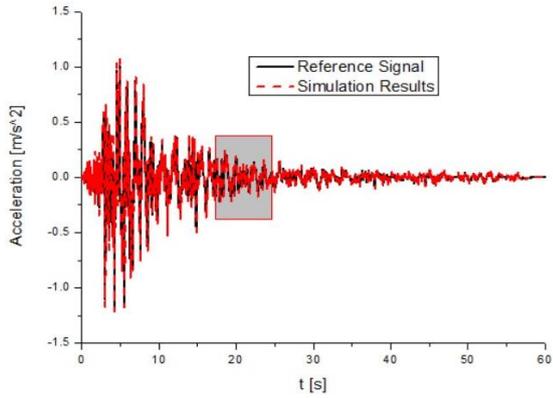
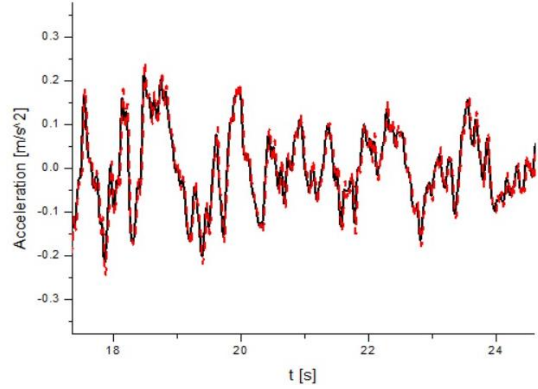

*Figure 4c Acceleration results of RSN 4013 from simulation*

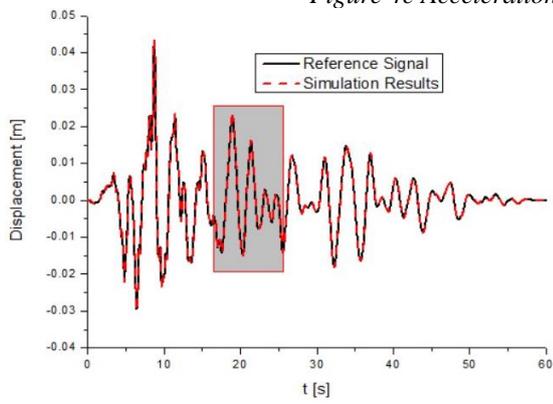
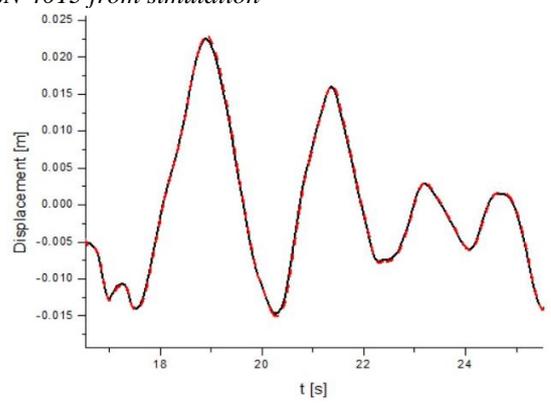

*Figure 5a Displacement results of RSN 1006 from simulation*

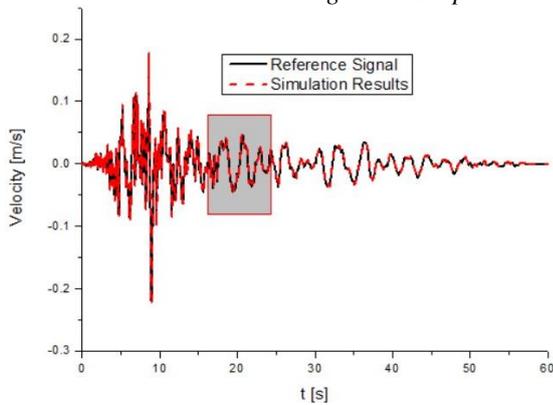
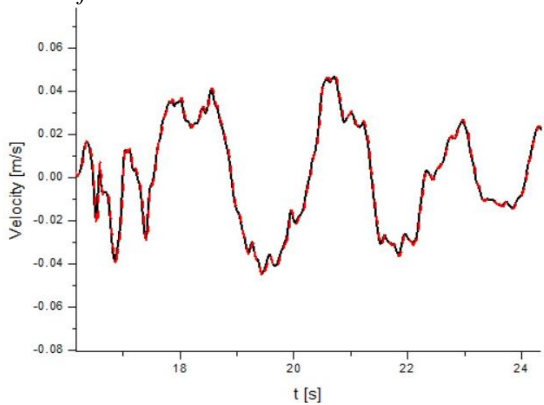

*Figure 5b Velocity results of RSN 1006 from simulation*

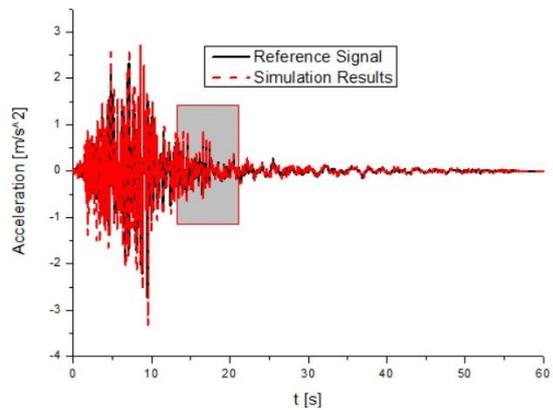
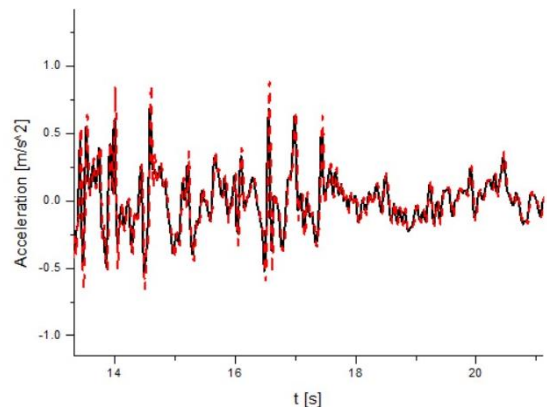

*Figure 5c Acceleration results of RSN 1006 from simulation*

## 3.3. Experimental results

The experiments were carried out in the Structural Engineering Laboratory at UBC. Figure (6) shows the experimental setup, which is a uniaxial shake table with a two-story frame. The testing results of the RSN 1006 are shown in Figure (7).

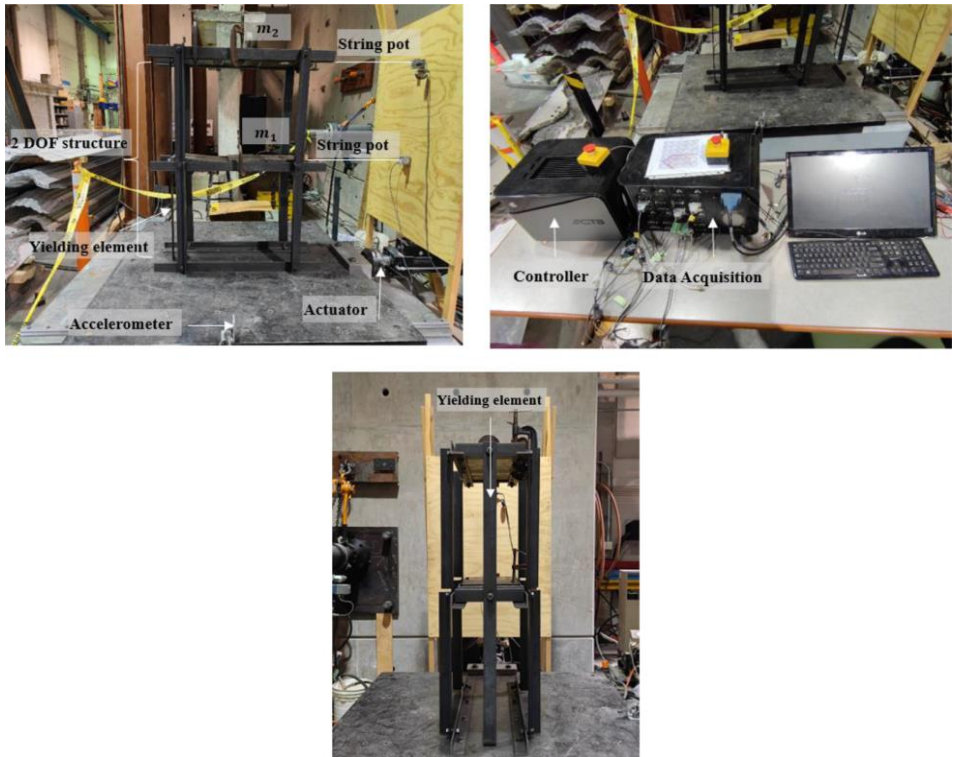

*Figure 6 Experimental setup [7]*

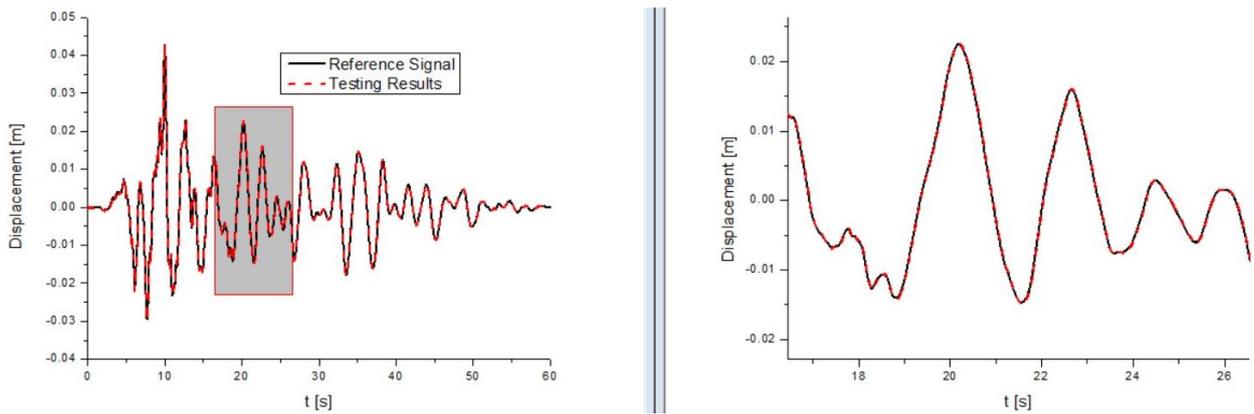

*Figure 7a Displacement results of RSN 1006 from testing*

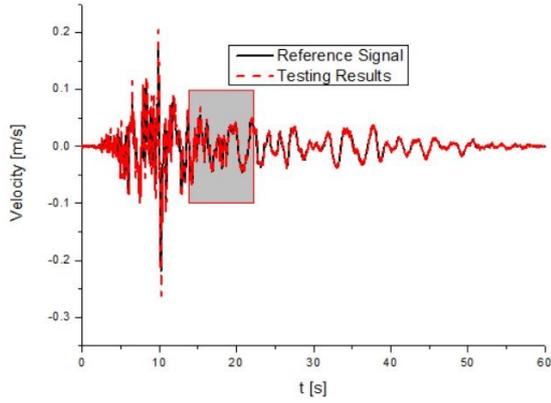 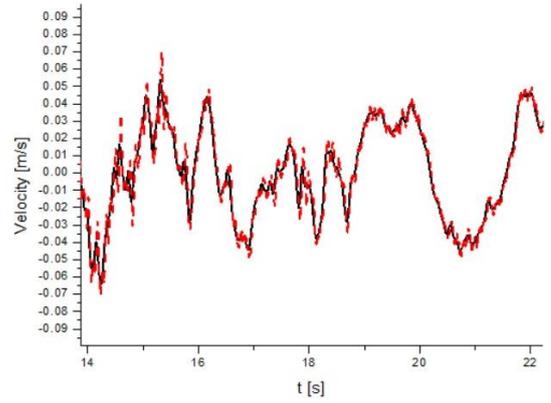

*Figure 7b Velocity results of RSN 1006 from testing*

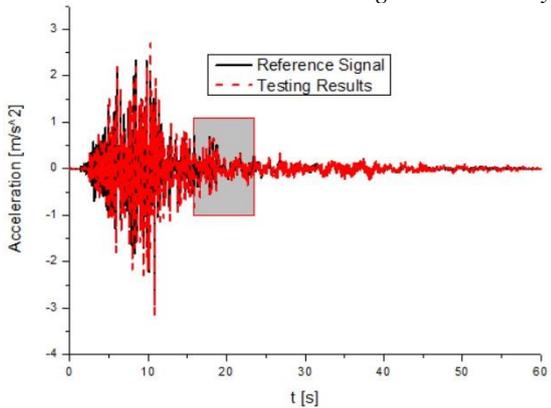 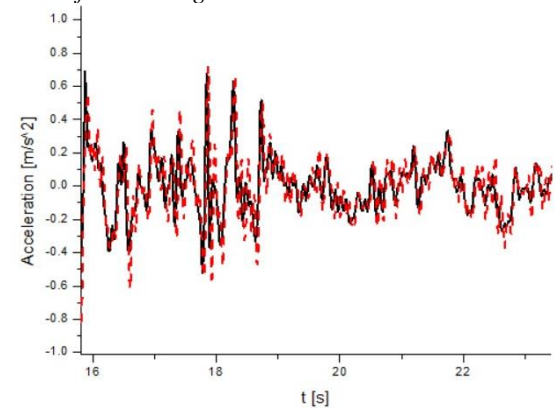

*Figure 7c Acceleration results of RSN 1006 from testing*

The normalized room mean square error (NRMSE) method is adopted to calculate the error between the reference signal and the measured output. The Equation of the NRMSE method is shown in Equation (17). The NRMSE of displacement is 0.7 %, the NRMSE of velocity is 3.6 %, and the NRMSE of acceleration is 10.3 %.

$$NRMSE = \frac{\sqrt{\frac{\sum_1^m (R_i - M_i)^2}{N}}}{\max(|R|)} \qquad (17)$$

where R and M are the reference signal and the measured signal respectively. N is the total number of signal points.
$i$ is $i^{th}$ signal point.

## 4. CONCLUSIONS

This paper proposed a nonlinear control algorithm to compensate for the CSI effect. The stability of the proposed nonlinear algorithm was valid by the Lyapunov theory first. Then the simulation and experiments were carried out to verify the stability and feasibility. The results present that the proposed control framework can be used in the nonlinear shake table control.

## 5. FUTURE STUDY

### 5.1. Comparison with classical control methods

The shake table test with different masses put on the upper structure will be carried out to verify the performance of the proposed control framework, and the results will be compared with the traditional control method, such as PID control and loop-shaping control. The displacement feedback of the shake table will be measured by computer vision.

## 5.2. Investigation of various computer vision methods in the control framework

As an active ongoing study, the authors are actively investigating the possibility of various computer vision algorithms to provide real-time sensor feedback for the proposed shake table control framework, in terms of the accuracy, speed, and robustness of various vision algorithms against various lighting conditions, as well as appropriate camera specs. Further, appropriate vision systems will be concluded for different testing setup, where a single camera can be used to capture the vibration response for multiple floors of the tested structures in real time and contactless manner. This will greatly reduce the total sensor cost and complexity of sensor installation compared to traditional shake table tests.